\documentclass[aps,prd,twocolumn,superscriptaddress,showpacs,amsmath,amssymb]{revtex4-1}
\usepackage{srcltx}
\newcommand{\be}{\begin{equation}}
\newcommand{\ee}{\end{equation}}
\newcommand{\bea}{\begin{eqnarray}}
\newcommand{\eea}{\end{eqnarray}}

\newcommand{\Z}{\mathbb{Z}}
\newcommand{\F}{\mathbb{F}}
\newcommand{\arbitraryext}{\,\ldotp}

\def\lsim{\mathrel{\rlap{\lower4pt\hbox{\hskip1pt$\sim$}}
    \raise1pt\hbox{$<$}}}         
\def\gsim{\mathrel{\rlap{\lower4pt\hbox{\hskip1pt$\sim$}}
    \raise1pt\hbox{$>$}}}         

\begin{document}

\textheight 24 cm

\title{Discrete symmetries in the three-Higgs-doublet model}

\author{Igor P. Ivanov}
\affiliation{IFPA, Universit\'{e} de Li\`{e}ge, All\'{e}e du 6 Ao\^{u}t 17, b\^{a}timent B5a, 4000 Li\`{e}ge, Belgium}
\affiliation{Sobolev Institute of Mathematics, Koptyug avenue 4, 630090, Novosibirsk, Russia}
\author{Evgeny Vdovin}
\affiliation{Sobolev Institute of Mathematics, Koptyug avenue 4, 630090, Novosibirsk, Russia}

\begin{abstract}
$N$-Higgs-doublet models (NHDM) are among the most popular examples of electroweak symmetry breaking mechanisms 
beyond the Standard Model.
Discrete symmetries imposed on the NHDM scalar potential play a pivotal role in shaping the phenomenology 
of the model, and various symmetry groups have been studied so far.
However, in spite of all efforts, the classification of finite Higgs-family symmetry 
groups realizable in NHDM for any $N>2$ is still missing.
Here, we solve this problem for the three-Higgs-doublet model
by making use of Burnside's theorem and other results from pure finite group theory
which are rarely exploited in physics.
Our method and results can be also used beyond high-energy physics, for example,
in study of possible symmetries in three-band superconductors.
\end{abstract}


\maketitle

\section{Introduction}

The nature of the electroweak symmetry breaking remains one of the hottest
issues in high-energy physics. 
The experimental quest for the Higgs boson, which was suggested back in 1964 \cite{higgs}, 
has very recently passed the first checkpoint: the CMS and ATLAS collaborations at the LHC
announced the discovery of the Higgs-like resonance at 126 GeV, \cite{discovery}.
Already their first measurements indicate intriguing
deviations from the Standard Model (SM) expectations.
Whether these data signal that a non-minimal Higgs mechanism is indeed at work 
and if so what it is, are among the hottest questions in particle physics these days.

Many different variants of non-minimal Higgs mechanism have been proposed so far \cite{CPNSh}.
One conceptually simple and phenomenologically attractive class of models involves 
several Higgs doublets 
with identical quantum numbers.
The scalar potential in these $N$-Higgs-doublet models (NHDM) is often assumed
to be symmetric under a group of unitary (Higgs-family) or anti-unitary (generalized $CP$)
transformations acting in the space of doublets.
These symmetries play a pivotal role in the phenomenology of the model, 
both in the scalar and in the fermionic sectors, \cite{flavour},
and they often bear interesting astrophysical consequences.
In fact, in many phenomenological models, one often starts by picking up a
symmetry group and then deriving phenomenological consequences.

In this situation, it is often very desirable to know which symmetry groups can be incorporated 
in a given model, and how they affect the phenomenological consequences.
Discrete symmetries are of special interest here due to a number of reasons.
First, unlike spontaneously broken continuous symmetries, they do not produce unwanted goldstone bosons. 
Second, finite symmetry groups with multi-dimensional irreducible representations often
lead to remarkable degeneracy patterns in the physical Higgs boson spectrum.
The simplest example here is an $S_3$-symmetric 3HDM with the 2HDM-like Higgs spectrum. 
Third, finite symmetry groups can lead to so-called geometric $CP$-violation, \cite{branco-gerard-grimus84,Delta27,geometricCP},
in which the calculable phases of vacuum expectation values are protected by the symmetry arguments.
Finally, there is a quest for derivation of the patterns observed in the fermion mixing matrices
from symmetry arguments, and finite groups are also at work here, \cite{flavour}.
Although these groups are introduced in the fermionic sector of the model, 
they might be related to symmetry groups in the Higgs sector, and the search for a convenient realization
of this link continues.

Given the high importance of symmetries for the NHDM phenomenology,
it is natural to ask: which symmetry groups can be implemented in the scalar sector of NHDM
for a given $N$?

In the two-Higgs-doublet model (2HDM), this question has been answered several years ago, 
\cite{2HDMsymmetries}, see also \cite{review2HDM} for a review.
Focusing on discrete symmetries, the only realizable group of unitary symmetries
are $\Z_2$ and $(\Z_2)^2$. If anti-unitary transformations are included, then $(\Z_2)^3$ is also realizable.
For each group, the corresponding potential was written and phenomenological consequences
were studied in detail (for example, an investigation of the $(\Z_2)^3$-symmetric 2HDM can be found see \cite{maximalCP}).

With more than two doublets, the problem remains open. 
Variants of NHDM based on several finite groups have been studied, \cite{variants}, with an emphasis on $A_4$,
\cite{A43HDM}, and $\Delta(27)$ or $\Delta(54)$, \cite{branco-gerard-grimus84,Delta27}.
Also, several attempts have been made to classify at least some symmetries in NHDM, 
\cite{symmetriesNHDM}. 
In particular, a classification of all realizable abelian symmetry groups in NHDM for any $N$ was recently given in
\cite{abelianNHDM}. However, the full list of non-abelian finite groups which can be symmetry groups in NHDM scalar sector
is not yet known.
We stress that this task is different from just classifying all finite subgroups of $SU(3)$, \cite{SU3finite}, 
because invariance of the Higgs potential places strong and non-trivial restrictions on possible symmetry groups.

In this paper we solve this problem for the three-Higgs-doublet model (3HDM). 
Starting from abelian groups and applying several results from
the finite group theory, we find the complete list of discrete symmetry groups of Higgs-family 
transformations realizable in 3HDM. 
Extension of our method to groups which include anti-unitary transformations
will be given elsewhere, \cite{nonabelian-future}.

We draw the reader's attention to the somewhat non-standard way we use group theory in our analysis.
Usually, group-theoretical methods in physics are limited to representation theory.
However, the formal group theory contains many powerful results beyond the representation theory 
which can be of great use for identifying symmetry properties of a model.
These results can restrict possible symmetries of the model without the need to explicitly manipulate 
with degrees of freedom which transform under specific representations of these groups.

This paper can be viewed as a non-trivial example of how powerful the pure group theory can be 
in finding symmetries of a model. Specifically, we make use of Burnside's theorem and other results
from finite group theory to solve the problem which seems to be out of reach for more traditional methods.
Although we focus below on this specific application, we stress that the same method can be also 
used in various condensed matter problems which involve several interacting order parameters \cite{GLmy}, 
in particular, in three-band superconductivity.

The structure of the paper is the following. In the next section we apply group-theoretic tools
to find the general structure of the finite groups which can be realized as 
Higgs-family symmetry groups in 3HDM. This result allows us to restrict the search for possible symmetry groups
to a very small set. Then, in section~\ref{section-construction}, we check all members of this set and see which 
groups can indeed be at work in 3HDM. We summarize our findings in section~\ref{section-summary}.

\section{Structure of finite symmetry groups in 3HDM}
The most general renormalizable gauge-invariant scalar potential of 3HDM can be written as
\be
\label{V:tensorial}
V = Y_{ij}(\phi^\dagger_i \phi_j) + Z_{ijkl}(\phi^\dagger_i \phi_j)(\phi^\dagger_k \phi_l)\,,
\ee
where all indices run from 1 to 3. We are interested in unitary transformations mixing 
doublets $\phi_i$ that leave this potential invariant for some $Y_{ij}$ and $Z_{ijkl}$.
A priori, these transformations belong to the group $U(3)$. Multiplying
the three doublets by a common phase factor, which trivially leaves the potential invariant, 
is already taken into account in the gauge group $U(1)_Y$. 
Therefore, we focus on additional transformations
not reducible to overall phase rotations, which form the group $PSU(3) = SU(3)/\Z_3$,
where $\Z_3$ is the center of $SU(3)$.
Our task is therefore to find finite subgroups of $PSU(3)$ which can be the symmetry groups
of the potential (\ref{V:tensorial}) for some choices of coefficients. 
We stress that we search for {\em realizable} symmetry groups, that is,
for groups $G \subset PSU(3)$ such that there exists a $G$-symmetric potential which
is not invariant under a larger symmetry group $G' \supset G$, see a fuller discussion in \cite{abelianNHDM}.

Abelian realizable symmetry groups for NHDM were characterized in \cite{abelianNHDM}. 
For our task of classifying finite realizable symmetry groups in 3HDM, the following abelian groups must be considered:
\be
\Z_2\,, \quad \Z_3\,, \quad \Z_4\,, \quad \Z_2\times \Z_2\,,\quad \Z_3 \times \Z_3\,.\label{abelian} 
\ee
The first four are the only realizable finite subgroups of maximal tori in $PSU(3)$.
The last group, $\Z_3 \times \Z_3$, is on its own a maximal abelian subgroup of $PSU(3)$,
but it is not realizable because a $\Z_3\times \Z_3$-symmetric potential is automatically symmetric
under $(\Z_3 \times \Z_3) \rtimes \Z_2$, see explicit expressions below. However, it still can appear as an abelian subgroup
of a finite non-abelian realizable group, therefore it must be included into consideration.
Trying to impose any other abelian Higgs-family symmetry group on the 3HDM potential 
unavoidably makes it symmetric under a continuous group.

Let us denote by $G \subset PSU(3)$ a finite (non-abelian) symmetry group in 3HDM.
We shall now apply some results from the finite group theory to prove that $G$ cannot be too large,
and more specifically, we shall describe the generic structure of $G$.

All abelian subgroups of $G$ must be from the list (\ref{abelian}).
By Chauchy's theorem, if $p$ is a prime divisor of the order of the group, $|G|$, then $G$ contains a subgroup $\Z_p$.
Thus, the order of the group can have only two prime
divisors: $|G| = 2^a 3^b$. Then according to Burnside's $p^aq^b$-theorem, the group $G$ is {\em solvable}.
Solvability implies that $G$ contains a normal abelian subgroup, which belongs, of course,
to the list (\ref{abelian}). This is our first key group-theoretic step.

Suppose $A$ is the normal abelian subgroup of $G$, $A \lhd G$.
Obviously, $A \subseteq C_G(A)$, the centralizer of $A$ in $G$ 
(all elements $g\in G$ which commute with all $a \in A$).
It turns out that this $A$ can be chosen in such a way that it coincides with its own
centralizer in $G$ (that is, it is elf-centralizing): $A = C_G(A)$, \cite{nonabelian-future}.
This means that elements $g \in G$, $g \not \in A$, cannot commute with {\em all} elements of $A$.
Therefore, they induce automorphisms (i.e. structure-preserving permutations) on $A$:  $g^{-1}ag \in A$ for any $a \in A$,
and these automorphisms are non-trivial. Even more, 
if $g_1$ and $g_2$ induce the same automorphism on $A$, $g_1^{-1}ag_1 = g_2^{-1}a g_2$ for all $a \in A$,
then $g_1$ and $g_2$ belong to the same coset of $A$ in $G$: $g_2 = g_1 a'$.
Therefore, the homomorphism
$f: G/A \to Aut(A)$, where $Aut(A)$ is the group of automorphisms on $A$, is {\em injective}.
We conclude that
\be
G/A = K\,,\quad K \subseteq Aut(A)\,.\label{GA}
\ee
This is our second key group-theoretic step. It proves that $G$ cannot be too large,
and it also shows that $G$ can be constructed as an extension of $A$ by a subgroup of $Aut(A)$:
$G = A \arbitraryext K$. 

\section{Explicit construction of possible symmetry groups}\label{section-construction}

We now check all the candidates for $A$ from the list (\ref{abelian})
and see which extension can work in 3HDM.
We use the explicit realization of each of the groups $A$, \cite{abelianNHDM}, 
and search for additional transformations from $PSU(3)$
with the desired multiplication properties.

\subsection{Extending $\Z_2$ and $\Z_3$}

If $A = \Z_2$, then $Aut(\Z_2) = \{1\}$, so that $G = \Z_2$.
This case was already considered in \cite{abelianNHDM}.

If $A = \Z_3$, then $Aut(\Z_3) = \Z_2$. The only non-trivial case to be considered is $G/A = \Z_2$, so that $G$ is
the dihedral group representing the symmetries of an equilateral triangle $G = D_6 = S_3$.
If $\Z_3$ group is generated by the phase rotations $a = \mathrm{diag}(\omega,\omega^2,1)$ with $\omega = 2\pi/3$,
then the transformation $b$ generating $\Z_2$ and satisfying $b^{-1}ab = a^2$
must be of the form
\be
b = \left(\begin{array}{ccc} 0 & e^{i\delta} & 0 \\ e^{-i\delta} & 0 & 0 \\ 0 & 0 & -1 \end{array}\right)\,,\label{bZ3}
\ee
with arbitrary $\delta$.
The choice of the mixing pair of doublets ($\phi_1$ and $\phi_2$ in this case) is also arbitrary,
so other $b$'s with different pairs of mixing doublets are also allowed.
The fact that $b$ is not uniquely defined means that there is a whole family of $D_6$ groups parametrized by the value of $\delta$
even if we start with the fixed group $A = \Z_3$.

The generic $\Z_3$-symmetric potential contains the part invariant under any phase rotation
\bea
V_0 &= & - \sum_{i} m_i^2(\phi_i^\dagger \phi_i) + \sum_{i,j} \lambda_{ij} (\phi_i^\dagger \phi_i)(\phi_j^\dagger \phi_j)\nonumber\\
&&+ \sum_{i \not = j} \lambda'_{ij} (\phi_i^\dagger \phi_j)(\phi_j^\dagger \phi_i)\,,\label{Tsymmetric}
\eea
and the following additional terms
\bea
V_{\Z_3} &=& \lambda_1(\phi_2^\dagger \phi_1)(\phi_3^\dagger \phi_1) + \lambda_2(\phi_1^\dagger \phi_2)(\phi_3^\dagger \phi_2)  
\nonumber\\
&&+ \lambda_3 (\phi_1^\dagger \phi_3)(\phi_2^\dagger \phi_3) + h.c.\label{VZ3}
\eea
with complex $\lambda_1,\, \lambda_2,\,\lambda_3$.
If the parameters of $V_0$ satisfy
\be
m_{11}^2 = m_{22}^2\,, \ \lambda_{11} = \lambda_{22}\,,\
\lambda_{13} = \lambda_{23}\,,\ \lambda'_{13} = \lambda'_{23}\,,\label{conditionV0}
\ee
and if, in addition, moduli of two among the three coefficients $\lambda_1$, $\lambda_2$, $\lambda_3$
coincide, for example $|\lambda_1| = |\lambda_2|$,
then the potential $V_0 + V_{\Z_3}$ becomes symmetric under one particular $D_6$ group
constructed with $b$ in (\ref{bZ3}) with the value of $\delta = (\arg \lambda_2 - \arg \lambda_1 + \pi)/3$.

This construction allows us to write down an example of the $D_6$ potential.
In order to prove that $D_6$ is indeed a realizable group, we need to show that 
the resulting potential is not symmetric under any other Higgs-family transformation.
This is proved by the mere observation that all other possible groups to be discussed below
which could contain $D_6$ lead to {\em stronger} restrictions on the potential than (\ref{conditionV0})
and $|\lambda_1| = |\lambda_2|$. 
Therefore, not satisfying those stronger restrictions
will yield a potential symmetric only under $D_6$.
Finally, one can also show that the potential we obtained does not have any continuous symmetry.
The same logic applies to other realizable groups below. 

\subsection{Extending $\Z_4$}

If $A = \Z_4$ (generated by $a$), then $Aut(\Z_4) = \Z_2$, so that $G = \Z_4\arbitraryext \Z_2$.
The two non-abelian possibilities for $G$ are
the dihedral group $D_8$ representing symmetries of the square, and the quaternion group $Q_8$. 
In both cases $b^{-1}ab = a^3$, with the only difference that $b^2 = 1$ for $D_8$ 
while $b^2 = a^2$ for $Q_8$.
Representing $a$ by $\mathrm{diag}(i,-i,1)$, 
we find
\be
b(D_8) = \left(\begin{array}{ccc} 0 & e^{i\delta} & 0 \\  e^{-i\delta} & 0 & 0 \\ 0 & 0 & -1 \end{array}\right)\!,
\nonumber\
b(Q_8) = \left(\begin{array}{ccc} 0 & e^{i\delta} & 0 \\  - e^{-i\delta} & 0 & 0 \\ 0 & 0 & 1 \end{array}\right).
\ee
Again, in each case we obtain a family of $b$'s parametrized by phase $\delta$.
The $\Z_4$-symmetric potential is $V_0 + V_{\Z_4}$ where
\be
V_{\Z_4} = \lambda_1 (\phi_3^\dagger \phi_1)(\phi_3^\dagger \phi_2) + \lambda_2 (\phi_1^\dagger \phi_2)^2 + h.c. \label{VZ4}
\ee
An explicit analysis shows that to make it $D_8$-invariant, we only need to satisfy conditions (\ref{conditionV0}).
Then, the potential is symmetric under $b(D_8)$ with the phase $\delta =  \arg \lambda_2/2$.
Since any larger group that could possibly contain $D_8$ leads to stronger restrictions
on the potential, we conclude that $D_8$ is realizable in 3HDM.

Now, if instead of $D_8$ we try to make the potential symmetric under $Q_8$, we unavoidably need to set $\lambda_1 = 0$.
Removing one term from (\ref{VZ4}) immediately makes it symmetric under a {\em continuous} group of phase rotations,
\cite{abelianNHDM}. Therefore, $Q_8$ is not realizable in 3HDM.

\subsection{Extending $\Z_2\times \Z_2$}

If $A = \Z_2 \times \Z_2$, then $Aut(\Z_2\times \Z_2) = GL_2(2) = S_3$. 
$\Z_2 \times \Z_2$ can be realized as the group of independent sign flips of the three doublets with
generators $a_1 = \mathrm{diag}(1,-1,-1)$ and $a_2 = \mathrm{diag}(-1,1,-1)$.
The potential symmetric under this group contains $V_0$ and additional terms
\be
V_{\Z_2 \times \Z_2} = \tilde \lambda_{12} (\phi_1^\dagger \phi_2)^2 +
\tilde \lambda_{23} (\phi_2^\dagger \phi_3)^2 +
\tilde \lambda_{31} (\phi_3^\dagger \phi_1)^2 + h.c.\label{VZ2Z2}
\ee
The coefficients $\tilde\lambda_{ij}$ can be complex; we denote their phases as $\psi_{ij}$.

There are three possibilities to extend $A$: by $\Z_2$, by $\Z_3$, and by $S_3$.
The first extension, $(\Z_2 \times \Z_2)\arbitraryext \Z_2$, leads to $D_8$, which was already constructed above.

The extension by $\Z_3$ is necessarily split, $(\Z_2 \times \Z_2)\rtimes \Z_3$, 
leading to the group $T \simeq A_4$, the symmetry group of a tetrahedron.
To construct it, we need to find $b$ acting on $\{a_1,a_2,a_1a_2\}$ by cyclic permutations.
Fixing the direction of permutations by $b^{-1}a_1 b = a_2$, we find that $b$ must be of the form
\be
b = \left(\begin{array}{ccc} 0 & e^{i\delta_1} & 0 \\ 0 & 0 & e^{i\delta_2} \\ e^{-i(\delta_1+\delta_2)} & 0 & 0 \end{array}\right)\,,
\ee
with arbitrary $\delta_1$, $\delta_2$.
It then follows that if coefficients in (\ref{VZ2Z2}) satisfy
\be
|\tilde \lambda_{12}| = |\tilde \lambda_{23}| = |\tilde \lambda_{31}|\,,\label{conditionT}
\ee
then $V_{\Z_2 \times \Z_2}$ is symmetric under a particular $b$ with
\be
\delta_1 = {2\psi_{12} - \psi_{31} - \psi_{23} \over 6}\,,\quad 
\delta_2 = {2\psi_{23} - \psi_{31} - \psi_{12} \over 6}\,.\nonumber
\ee
By rephasing, one can bring (\ref{VZ2Z2}) to the following form
\be
V_T = \tilde\lambda  \left[(\phi_1^\dagger\phi_2)^2 + (\phi_2^\dagger\phi_3)^2 + (\phi_3^\dagger\phi_1)^2\right] + h.c.\label{VT}
\ee
with complex $\tilde \lambda$.
In addition, the symmetry under $b$ places stronger conditions on the parameters of $V_0$, and the most general $V_0$
satisfying them is now 
\bea
\label{V0Z2Z2Z3}
V_0 &=& -m^2 \sum_{i} (\phi_i^\dagger\phi_i) + 
\lambda \left[\sum_{i} (\phi_i^\dagger\phi_i)\right]^2  \\
&&+ \sum_{i\not = j}\left[\lambda'(\phi_i^\dagger\phi_i)(\phi_j^\dagger\phi_j)
+ \lambda'' |\phi_i^\dagger\phi_j|^2\right]\,.\nonumber
\eea
The last extension, $(\Z_2 \times \Z_2)\arbitraryext S_3$, leads to the group $O=S_4$, the symmetry group of an
octahedron and a cube.
It includes $T$ as a subgroup, therefore the most general $O$-symmetric potential is $V_0$ from (\ref{V0Z2Z2Z3})
plus $V_T$ from (\ref{VT}) with the additional condition that $\tilde \lambda$ is real.

\subsection{Extending $\Z_3\times \Z_3$}

Finally, if $A = \Z_3\times \Z_3$, then $K \subset Aut(\Z_3\times \Z_3) = GL_2(3)$,
the general linear group of transformations of two-dimensional vector space over the finite field $\F_3$,
whose role is played by $A$.
One can define an antisymmetric scalar product in this space and prove that $K$ must include only 
transformations from $GL_2(3)$ that preserve this scalar product:
$K \subseteq Sp_2(3) = SL_2(3)$. 

The group $SL_2(3)$ has order 24 and contains
elements of order 2, 3, 4, and 6. Elements of order 6 cannot be used for extension because they would
generate the abelian subgroup $\Z_6$, which is absent in (\ref{abelian}).
Besides, we will show below that $K$ must always contain the subgroup $\Z_2$.
There are three kinds of subgroups of $K \subset SL_2(3)$ containing $\Z_2$ but not containing $\Z_6$:
$\Z_2$, $\Z_4$, and $Q_8$. Since, as we argued, the quaternion group $Q_8$ is not realizable
in 3HDM, $K$ can only be $\Z_2$ or $\Z_4$.

To show that $K \supseteq \Z_2$, consider first the subgroup of $SU(3)$ generated by  
\be
a = \left(\begin{array}{ccc} 1 & 0 & 0 \\ 0 & \omega & 0 \\ 0 & 0 & \omega^2 \end{array}\right),\quad
b = \left(\begin{array}{ccc} 0 & 1 & 0 \\ 0 & 0 & 1 \\ 1 & 0 & 0 \end{array}\right),\quad \omega = \exp\left({2\pi i \over 3}\right)\,.
\nonumber
\ee
This subgroup is usually denoted as $\Delta(27)$, \cite{fairbairn}, and it is well-known in model-building with three Higgs doublets,
\cite{branco-gerard-grimus84}. 
Since $[a,b]=aba^{-1}b^{-1} \in Z(SU(3))$, the image of $\Delta(27)$ under the canonical homomorphism
$SU(3) \to PSU(3)$ becomes the desired abelian group $\Z_3 \times \Z_3$. 
The true generators of $\Z_3 \times \Z_3$ are cosets $\bar a = aZ(SU(3))$ and $\bar b = bZ(SU(3))$ from $PSU(3)$.
The $\Z_3\times \Z_3$-invariant potential is
\begin{widetext}
\bea
V & = &  - m^2 \left[(\phi_1^\dagger \phi_1)+ (\phi_2^\dagger \phi_2)+(\phi_3^\dagger \phi_3)\right]
+ \lambda_0 \left[(\phi_1^\dagger \phi_1)+ (\phi_2^\dagger \phi_2)+(\phi_3^\dagger \phi_3)\right]^2 \nonumber\\
&&+ {\lambda_1 \over \sqrt{3}} \left[(\phi_1^\dagger \phi_1)^2+ (\phi_2^\dagger \phi_2)^2+(\phi_3^\dagger \phi_3)^2
- (\phi_1^\dagger \phi_1)(\phi_2^\dagger \phi_2) - (\phi_2^\dagger \phi_2)(\phi_3^\dagger \phi_3) 
- (\phi_3^\dagger \phi_3)(\phi_1^\dagger \phi_1)\right]\nonumber\\
&&+ \lambda_2 \left[|\phi_1^\dagger \phi_2|^2 + |\phi_2^\dagger \phi_3|^2 + |\phi_3^\dagger \phi_1|^2\right] 
+ \lambda_3 \left[(\phi_1^\dagger \phi_2)(\phi_1^\dagger \phi_3) + (\phi_2^\dagger \phi_3)(\phi_2^\dagger \phi_1) + (\phi_3^\dagger \phi_1)(\phi_3^\dagger \phi_2)\right]
+ h.c.\label{VZ3Z3}
\eea
\end{widetext}
with real $m^2$, $\lambda_0$, $\lambda_1$, $\lambda_2$ and complex $\lambda_3$, all values being generic.
This potential is, however, symmetric under a larger group $(\Z_3\times \Z_3)\rtimes \Z_2 \simeq \Delta(54)/\Z_3$,
which is generated by $\bar a,\bar b,\bar c$ with the following relations
\be
\bar a^3 = \bar b^3 = 1,\ \bar c^2 = 1,\ [\bar a,\bar b]=1,\ \bar c\bar a\bar c^{-1} = \bar a^2,\ \bar c\bar b\bar c^{-1} = \bar b^2\,.\nonumber
\ee
In terms of the explicit transformation laws, $\bar c$ is the coset $cZ(SU(3))$, with $c$ being the exchange of any two doublets,
so that $\langle \bar a, \bar c\rangle = S_3$ is the group of arbitrary permutations of the three doublets.
Thus, if $G = (\Z_3\times \Z_3)\arbitraryext K$, then a $G$-symmetric potential must be a restriction of (\ref{VZ3Z3}), 
so that $K \supseteq \Z_2$. 

Turning now to the extension $G = (\Z_3\times \Z_3)\rtimes \Z_4$ (which is also known as $\Sigma(36)$, \cite{fairbairn}),
we note that $SL_2(3)$ contains three distinct $\Z_4$ subgroups, which however intersect at the center of $SL_2(3)$. 
All three are conjugate inside $SL_2(3)$ and lead, up to isomorphism, to the same symmetry group.
To give an example of a potential symmetric under $(\Z_3\times \Z_3)\rtimes \Z_4$,
we choose an element $d \in SL_2(3)$ of order 4 that generates the cyclic permutation of generators $\bar a, \bar b, \bar a^2, \bar b^2$.
It can be represented by the following $SU(3)$ transformation:
\be
d = {i \over\sqrt{3}} \left(\begin{array}{ccc} 1 & 1 & 1 \\ 1 & \omega^2 & \omega \\ 1 & \omega & \omega^2 \end{array}\right)\,,\quad
d^{-1} = d^*\,, \quad d^4=1\,.
\ee
Then by analyzing how the potential changes under $d$, we obtain the following criterion:
(\ref{VZ3Z3}) becomes symmetric under $(\Z_3\times \Z_3)\rtimes \Z_4$, if 
$\lambda_3$ is real and is equal to $(\sqrt{3}\lambda_{1} - \lambda_2)/2$.
One can also show that the resulting potential is not invariant under any continuous symmetry group.

\section{Summary}\label{section-summary}

Because of important phenomenological role the symmetries play in multi-Higgs-doublet models,
the task of classifying all symmetries in NHDM is of much interest.
Here we solved this problem for 3HDM. Focusing on groups of unitary
transformations and including the finite abelian groups found in \cite{abelianNHDM}, 
we obtain the following list of finite groups realizable as Higgs-family symmetry groups 
of the 3HDM scalar sector:
\bea
&&\Z_2\,, \quad \Z_3\,, \quad \Z_4\,, \quad \Z_2\times \Z_2\,,\\ 
&&D_6\,,\quad D_8\,, \quad T \simeq A_4\,,\quad O \simeq S_4\,,\nonumber\\ 
&&(\Z_3 \times \Z_3)\rtimes \Z_2\simeq \Delta(54)/\Z_3\,, \quad \ (\Z_3 \times \Z_3)\rtimes \Z_4\simeq \Sigma(36)\,.\nonumber
\eea
This list is complete:
trying to impose any other finite Higgs-family symmetry group on the 3HDM potential
will unavoidably lead to a potential symmetric under a continuous group.
Applying methods described in \cite{abelianNHDM}, 
one can also obtain 
the list of realizable groups in 3HDM which include
anti-unitary transformations. These results as well as a study of symmetry breaking patterns
for each of these groups will be presented elsewhere, \cite{nonabelian-future}.

We stress that we solved the classification problem in a rather non-standard way, by applying results and tools
from formal finite group theory and without using representation theory.
We view this as an example of how powerful pure group theory can be in identifying symmetries of
a model. We also stress that the same method can be applied to problems beyond 
particle physics, for example, for understanding symmetries of the order parameters in three-band superconductors.

\section*{Acknowledgements}
This work was supported by the Belgian Fund F.R.S.-FNRS, 
and in part by grants RFBR 11-02-00242-a, RFBR 10-01-0391, RF President grant for
scientific schools NSc-3802.2012.2, and the 
Program of Department of Physics SC RAS and SB RAS "Studies of Higgs boson and exotic particles at LHC."

\end{document}